\documentclass[superscriptaddress, preprint, amsmath, amssymb, aps, physrev]{revtex4-2}

\usepackage{siunitx}
\usepackage[english]{babel} 

\usepackage[utf8]{inputenc}
\DeclareUnicodeCharacter{2212}{-}
\usepackage{chemformula}

\begin{document}
\title{ $\beta$-\ch{Ga2O3}-Based Heterojunctions: Effects of Growth Orientation and Alloying on Electronic Properties
}


\author{Mohamed Abdelilah Fadla}
\email{m.fadla@qub.ac.uk}
\affiliation{School of Mathematics and Physics, Queen's University Belfast, University Road, Belfast BT7 1NN, UK}

\author{Khushabu Agrawal }
\affiliation{Tyndall National Institute, University College Cork, Lee Maltings, Prospect Row, Cork, Ireland}
\author{Paolo La Torraca}
\affiliation{Tyndall National Institute, University College Cork, Lee Maltings, Prospect Row, Cork, Ireland}
\author{Myrta Gr\"{u}ning}
\affiliation{European Theoretical Spectroscopy Facility}
\affiliation{School of Mathematics and Physics, Queen's University Belfast, University Road, Belfast BT7 1NN, UK}
\author{Karim Cherkaoui}
\affiliation{Tyndall National Institute, University College Cork, Lee Maltings, Prospect Row, Cork, Ireland}
\author{Lorenzo Stella}
\affiliation{School of Mathematics and Physics, Queen's University Belfast, University Road, Belfast BT7 1NN, UK}


\begin{abstract}
We investigate the effects of alloying and growth orientation on the electronic properties of the ultra-wide bandgap semiconductor $\beta$-\ch{Ga2O3} and pseudomorphic \ch{(Al_xGa_{1-x})2O3} alloy heterojunctions. Band offsets are computed from first principles using density functional theory (DFT) with the Heyd-Scuseria-Ernzerhof hybrid functional for different Al concentrations and four growth orientations, namely (\(100\))B, (\(010\)), (\(001\))B, and  (\(\bar{2}01\)). Significant variations are found and ascribed to the strained pseudomorphic alloys. The values of the band offsets are fed into technology computer-aided design (TCAD) models of Schottky barrier diodes (SBD).  I--V and C--V characteristics from the TCAD models show reasonable agreement with recent experimental measurements in the forward bias region. Discrepancies in the negative bias region are expected due to the ideality of the Schottky junctions considered in this study. Our findings underscore the critical role of growth orientation and strain in the accurate modelling of $\beta$-\ch{Ga2O3}-based SBD.
\end{abstract}

\maketitle
\section{Introduction}

\ch{Ga2O3}, with its wide bandgap of approximately 4.7 eV, has garnered significant attention as a promising material for various applications, including metal oxide-semiconductor field-effect transistors (MOSFETs), UV solar-blind photodetectors, and Schottky barrier diodes (SBDs) \cite{higashiwaki_gallium_2012,hwang_high-voltage_2014}. \ch{Al2O3} is frequently considered as gate oxide in \ch{Ga2O3}-based heterojunctions. \cite{krueger_variation_2016, li_identification_2018, peelaers2018structural, ratnaparkhe_quasiparticle_2020, mu_first-principles_2020, kim2021heterostructural} Differences in the crystal structures of \ch{Ga2O3} and \ch{Al2O3} can lead to lattice mismatches in heterojunctions. Cracking may occur for film thicknesses exceeding a critical value. \cite{mu_first-principles_2020, fadla2024effective} To mitigate these mismatches --- and also to engineer electronic properties such as the band gap --- alloying \ch{Ga2O3} with aluminium \ch{(Al_xGa_{1-x})2O3} is widely employed. \cite{krueger2016variation,kranert2015lattice,wang2016temperature,li_identification_2018,peelaers2018structural,ratnaparkhe_quasiparticle_2020,mu_first-principles_2020,kim2021heterostructural} 
In applications such as SBD and high electron mobility transistors, band offset is crucial in developing semiconductor devices. The reported values of CBO can vary significantly, by up to 1.5 eV. \cite{lyu2023band,wang2018band,hinuma2019band,seacat2024computational,hinuma2019band,seacat2020orthorhombic,wang2018band,mu2020orientation,kim2021heterostructural,peelaers2018structural,lyu2023band,carey2017band,peelaers2018structural,seacat2024computational,hinuma2019band,mu2020orientation,kim2021heterostructural,kamimura2014band,hung2014energy,hattori2016epitaxial} 
Experimental studies, such as those using capacitance-voltage (C--V) profiling and X-ray photoelectron spectroscopy (XPS) 
exhibit strong variation, which also depends on the methods used for deposition (\emph{e.g.}, sputtering-prepared films, pulsed laser deposition, atomic layer deposition) and the characterisation techniques employed.\cite{carey2017band,kamimura2014band,hung2014energy,hattori2016epitaxial}  Theoretical studies, typically based on atomistic modeling, provide subresolution information regarding the large anisotropy of these alloys, where the monoclinic phase’s low symmetry results in varying surface growth orientations. Most of the atomistic modelling reported in the literature considered isolated surfaces \emph{i.e.}, offset computed using the electron affinity rule. \cite{peelaers2018structural,seacat2024computational,seacat2020orthorhombic,kim2021heterostructural} Although these models included surface relaxation, only a few studies considered proper interfaces to model these heterojunctions.\cite{hinuma2019band,wang2018band,mu2020orientation,lyu2023band} We note here that, while most of these reports investigate how the band offset and electron affinity are affected by both composition and structure, in nearly all cases, only the natural band offset is considered. 
In these cases, the effect of strain, as well as the chemical bonding at the interfaces, is neglected, which can lead to inaccurate predictions of both the magnitude and the nature of the band offset.\cite{lyu2023band,mu2020orientation}

Here, using density functional theory (DFT) at the hybrid functional level, we investigate the $\beta$-\ch{Ga2O3}/\ch{(Al_xGa_{1-x})2O3} 
interface.  The accurate prediction of the band offset and its dependence on the crystallographic orientation within the material is crucial to improve the overall performance of devices. The monoclinic phase of \ch{Ga2O3} is considered as the substrate, assuming that the grown film has the same structure and orientation as the substrate, i.e., pseudomorphic growth. The calculated band offsets and affinities are integrated into technology computer-aided design (TCAD) models of SBD to study their electrical characteristics. In the following, we initially provide a detailed description of the modelling methodology, followed by the presentation of our findings regarding the electronic parameters of substrates and films, considering strain effects. Then, we explore the band offsets of pseudomorphically grown films \ch{(Al_xGa_{1-x})2O3} 
, exploring their significance for device design. We found that orientation and strain have a critical role in the accurate modelling of these heterojunctions and thus in optimising wide-bandgap devices for power electronics.

\section{Computational details}

The electronic structure was calculated using density functional theory (DFT) with the Vienna ab initio simulation package (VASP).\cite{kresse_efficient_1996,kresse1996efficiency,kresse1993ab} For structural properties, full optimisation, the strongly constrained and appropriately normed (SCAN)\cite{sun_strongly_2015} functional was initially used. Additionally, for atomic relaxation and electronic properties, the Heyd-Scuseria-Ernzerhof (HSE06)\cite{10.1063/1.1564060,krukau_influence_2006} functional was used with a mixing parameter of 0.32. A gamma sampling scheme for the Brillouin zone was used, with a 0.03 (0.025) 1/Å separation for structural (electronic) calculations and 600 eV as the plane wave cutoff energy. The projected augmented wave (PAW) pseudopotential \cite{blochl_projector_1994} was employed with the following valence configurations:  \textit{d}$^{10}$\textit{s}$^{2}$\textit{p}$^{1}$ for gallium, \textit{s}$^{2}$\textit{p}$^{1}$ for aluminium, and \textit{s}$^{2}$\textit{p}$^{4}$ for oxygen. 

The structures were relaxed until the maximum force components reached a value less than 0.01 eV/Å, with an energy convergence threshold set to $10^{-6}$ eV for all supercell calculations.
Both $\theta$-\ch{Al2O3} and $\beta$-\ch{Ga2O3} crystallize in the monoclinic space group C2/m, with 20-atom conventional unit cells \cite{ahman_reinvestigation_1996}. In both compounds, trivalent Al or Ga cations occupy octahedral and tetrahedral sites.  
The interfaces were constructed using a periodic supercell along the out-of-plane direction, with no vacuum. We refer to this construction as the superlattice structures. The in-plane lattice constants of the films are fixed to those of the substrate. 

Different orientations and the most stable terminations were considered, namely (100), (010), (001), ($\bar{2}$01). In the cases of (100) and (001) orientations, there are two possible terminations. The most stable are (100)B and (001)B. (Superlattice structures are shown in the supporting information). Only symmetric and non-polar terminations were considered. The thickness of each layer of the superlattice structures (film: \ch{Al2O3}, and substrate: \ch{Ga2O3}) was set to ensure that the electron density in the middle of the layer is the same as in the bulk reference.

The average electrostatic (\emph{i.e.}, Hartree plus external) potential is used as a reference for bulk calculation and to compute the potential lineup between both sides of the interface. The valence band offset (VBO) was computed as
\begin{equation}
\text{VBO}(A/B) = E_{\text{VBM,Ref}}^B - E_{\text{VBM,Ref}}^A + \Delta \overline{V}^{AB},
\label{eq:eq1}
\end{equation}
where $E_{\text{VBM,Ref}}^B$ and $E_{\text{VBM,Ref}}^A$ are the energies of the valence band maximum of bulk materials B and A, respectively, measured relative to a common reference level, such as the average electrostatic potential. $\Delta \overline{V}^{AB}$ is the average electrostatic potential difference between the two sides of the interface.

The CBO was calculated using the same approach as
\begin{equation}
\text{CBO}(A/B) =  E_{\text{CBM,Ref}}^B - E_{\text{CBM,Ref}}^A + \Delta \overline{V}^{AB},
\label{eq:eq2}
\end{equation}
where $E_{\text{CBM,Ref}}^B$ and $E_{\text{CBM,Ref}}^A$ are the energies of the conduction band minimum of bulk materials B and A, respectively.

To investigate the dependence of the band gap on the alloy composition, 1$\times$3$\times$2 special quasi-random structure (SQS) supercells were employed. Two- and three-atom clusters are used to minimize the correlation mismatch between SQS and truly random alloy structures, as implemented in the \texttt{mcsqs} code distributed with the Alloy Theoretic Automated Toolkit (ATAT) \cite{van_de_walle_efficient_2013}. Al cations preferentially occupy octahedral sites\cite{mu_first-principles_2020,ratnaparkhe_quasiparticle_2020}. Accordingly, for the generation of SQS at Al concentrations below 50 \%  (\textit{i.e.}, $x<0.5$), Al cations are randomly substituted among the octahedral sites. For Al concentration exceeding 50 \%, all octahedral sites become fully occupied, and the remaining Al cations are randomly substituted into the tetrahedral sites.
For \ch{(Al_{0.5}Ga_{0.5})2O3} film, an additional superlattice structure was employed in which all octahedral sites of the film are occupied by Al cations. The values of the average electrostatic potential differences, $\Delta \bar{V}^{AB}$, for \ch{(Al_xGa_{1-x})2O3} alloys are obtained by means of linear interpolation.

$\beta$-\ch{(Al_xGa_{1-x})2O3}/Ga$_2$O$_3$ device simulations were performed using Applied Materials' Ginestra\textregistered\  ,\cite{appliedmaterials2025} 
a state-of-the-art simulation software which provides a comprehensive description of the transport in electronic devices. In Ginestra, all the relevant charge transport mechanisms are simultaneously accounted (including drift-diffusion in the materials’ bands, thermionic emission, direct tunnelling, Fowler-Nordheim tunnelling, and multiphonon trap-assisted tunnelling), with the relative equations self-consistently solved together with Poisson’s equation.
The simulation activity focused on generating the ideal current-voltage (I--V) and capacitance-voltage (C--V) characteristics of Pt/$\beta$-\ch{(Al_xGa_{1-x})2O3}/Ga$_2$O$_3$  SBD, considering the parameters provided by the ab-initio DFT calculations presented in this study. Comparisons with experimental I--Vs previously published in literature are also presented, validating the presented DFT calculations results. The main parameters used for the simulations are listed in Table S1 \cite{p_sundaram_-alxga1x2o3ga2o3_2022,guo_review_2019,mudiyanselage_ultrawide_2023}.

\section{Results and discussion}
\subsection{Pseudomorphic \ch{Al2O3} films on $\beta$-\ch{Ga2O3} substrates}

Due to the lattice mismatch between \ch{Al2O3} films and the \ch{Ga2O3} substrates, an in-plane tensile strain is induced during the \ch{Al2O3}  growth. For each pseudomorphic growth direction --- namely (100)B, (010),  (001)B, and ($\bar{2}01$) --- the in-plane lattice parameters of the \ch{Al2O3} films were matched to those of the $\beta$-\ch{Ga2O3} substrate, while the structures were relaxed along out-of-plane direction. We refer to these strained \ch{Al2O3} structures as the ``films". The in-plane tensile strain of the films causes significant shifts in the \ch{Al2O3} band structure. The magnitude of these shifts can be evaluated from the band gaps, valence band maximum (VBM), and conduction band minimum (CBM) as reported in Table \ref{Table1}. In the same Table,  the lattice parameters for bulk $\beta$-\ch{Ga2O3}, bulk  $\theta$-\ch{Al2O3}, and \ch{Al2O3} films are also reported. 
\begin{table}[htbp]
\centering
\begin{tabular}{|c|c|c|c|c|c|c|}
\hline
 & $\beta$-\ch{Ga2O3}& $\theta$-\ch{Al2O3}& (100)B & (010) & (001)B & ($\bar{2}01$) \\ 
\hline
a& 12.21& 11.77& 3.04& 5.83 & 12.21& 3.04\\ \hline 
 b& 3.04& 2.90& 5.83 & 12.21& 3.04&14.71\\
\hline
c & 5.83 & 5.61 & 11.31 & 2.87 & 5.52 & 5.53 \\\hline
 $V_{\text{f.u.}} $& 52.54& 46.48& 48.64& 49.47& 49.75&49.95\\\hline 

E\(_g\) & 4.71 & 7.19 & 6.62 & 6.49 & 6.51 & 6.45 \\ 
\hline
E$_{\text{CBM,Ref}}$ & 6.875 & 11.022 & 9.997 & 9.634 & 9.651 & 9.580 \\ 
\hline
E$_{\text{VBM,Ref}}$ & 2.17 & 3.832 & 3.378 & 3.140 & 3.146 & 3.133 \\ 
\hline
\end{tabular}
\caption{Calculated lattice parameters (a, b, c) and volume per formula unit ($V_{\text{f.u.}} $) obtained using the SCAN functional. The band gaps (E$_g$) and relative positions of the E$_{\text{CBM,Ref}}$ and E$_{\text{VBM,Ref}}$ are determined using the HSE06 hybrid functional for Ga$_2$O$_3$, unstrained bulk $\theta$-\ch{Al2O3} and strained $\theta$-\ch{Al2O3} bulk corresponding to the most common  crystal growth orientations. The strained in-plane lattice constants are denoted as a and b, while the relaxed out-of-plane constant is c. The transformation matrix for ($\bar{2}01$) orientation is provided in the Supporting Information.}
\label{Table1}
\end{table}

Note that the CBM and VBM energy reported in Table \ref{Table1} are those used in Eqs. \ref{eq:eq1} and \ref{eq:eq2} as reference energy. That is because the films structures have been relaxed independently, \emph{i.e.}, before forming an interface with the $\beta$-\ch{Ga2O3} substrate. On the other hand, the average electrostatic potential difference,  $\Delta \overline{V}^{AB}$, has been computed after the interface has been formed, \emph{i.e.,} after full relaxation of the heterostructure. 
It is evident from the energies reported in Table \ref{Table1} that the strain yields a notable effect on the band gaps of the \ch{Al2O3} films, which is smaller than that of bulk $\theta$-\ch{Al2O3}. This is also true for the CBM and VBM of the films, although the variation of the VBM is less pronounced. Given these sizeable variations of the reference energies, it is crucial to relax the film structure before computing the CBO and VBO using Eqs.  \ref{eq:eq1} and \ref{eq:eq2}. Among the studied orientations, (100)B exhibits the smallest band-edge shifts, originating from its lower net strain (and thus smaller volume change, see Table \ref{Table1}) compared to other orientations. As a consequence, the change in band gap is also minimal.

The role of the strain in determining the band offset in heterostructures is also shown in Ref.~\cite{lyu2023band}. There, two different substrate--film combinations were used: Ga$_2$O$_3$ as the substrate with an Al$_2$O$_3$ film or Al$_2$O$_3$ as substrate with a Ga$_2$O$_3$ film. In both cases, the films' in-plane lattice parameters were constrained to match those of the respective substrate lattice. Results of Ref.~\cite{lyu2023band} show that swapping substrates with films not only affects the magnitude of the band offset, but also changes the type of band offset, as the VBO changes sign, from -0.14 eV (when $\beta$-\ch{Ga2O3} is the substrate) to 0.54 eV (when \ch{Al2O3} is the substrate).

Figure S2 illustrates the $\theta$-\ch{Al2O3}/$\beta$-\ch{Ga2O3} superlattice structures for the (100)B, (010), (001)B, and ($\bar{2}01$) orientations. As previously discussed, only nonpolar, symmetric and most stable terminations were employed to prevent any built-in potential difference at the interfaces. The in-plane lattice parameters were fixed to those of the substrate, while the out-of-plane lattice parameter of the superlattice structures were relaxed to release the strain along that direction. After the cell relaxations,  full atomic relaxation at fixed cell structure was performed.  

\begin{figure}[h]
\includegraphics[width=0.65\textwidth]{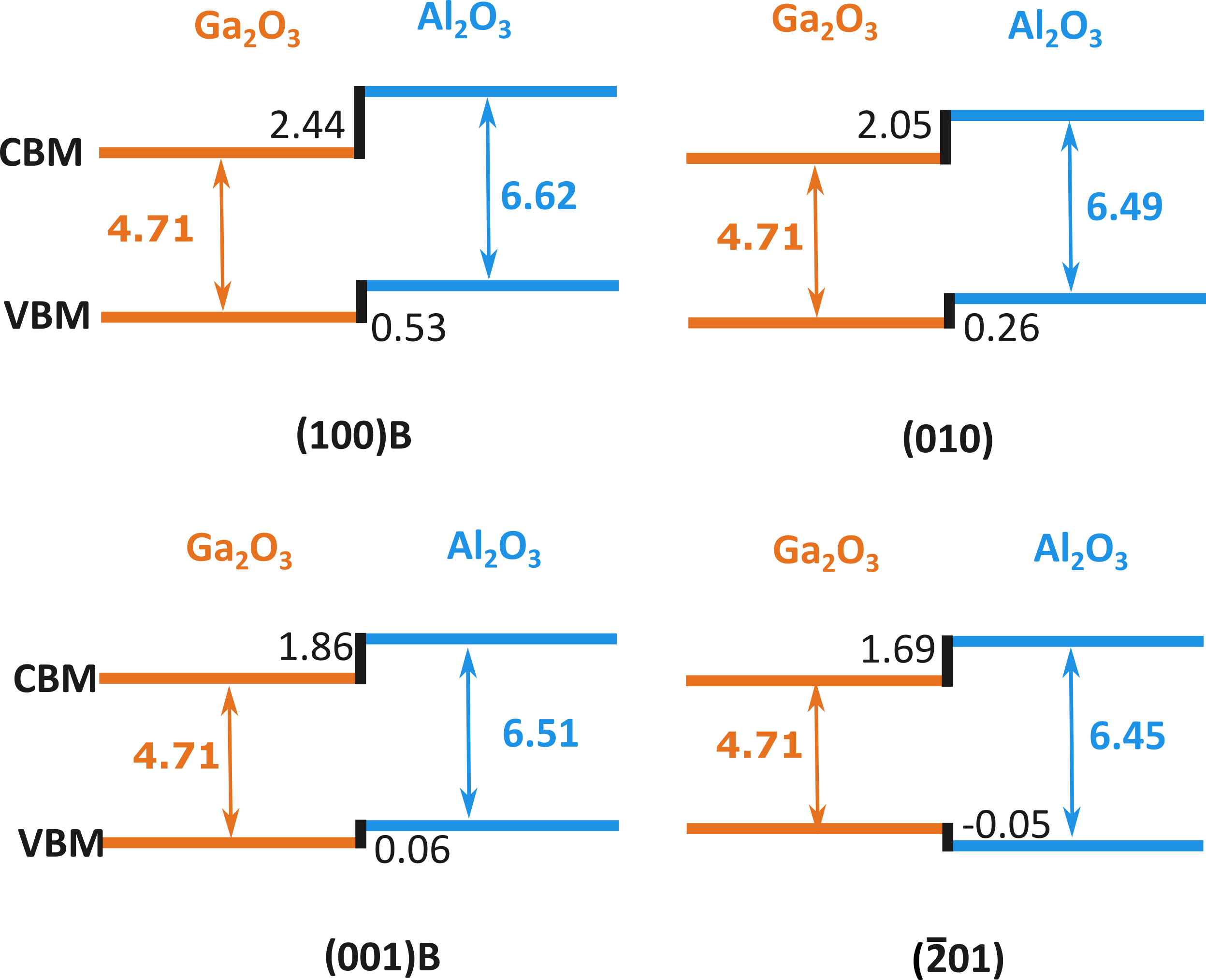}
\caption{Calculated band offsets (in eV) between $\beta$-\ch{Ga2O3} and $\theta$-\ch{Al2O3} for (100)B, (010), (001)B, and ($\bar{2}$01) orientations.}
\label{fig:fig1}
\end{figure}

The averaged electrostatic potential profiles are presented in Figure S4. The planar average electrostatic potential is computed by averaging the electrostatic potential over the in-plane directions, while the macroscopic average is computed along the out-of-plane direction. 
An essentially constant macroscopic average indicates a bulk-like behaviour, confirming the absence of any built-in potential at the interfaces. A constant macroscopic average potential serves as a reference for the band offset calculations. The alignment procedure described in Eqs.~(\ref{eq:eq1})  and (\ref{eq:eq2}) is then applied to obtain the band offset.  
Figure~\ref{fig:fig1} illustrates the VBO  and CBO of the \ch{Ga2O3}/\ch{Al2O3} heterojunctions, highlighting their dependence on different interface orientations, namely ((100)B, (010), (001)B, and ($\bar{2}01$)).

The calculated values vary significantly across orientations, not only in magnitude but also in type. For instance, the ($\bar{2}01$) orientation exhibits a band offset (type I straddling) different from that of the other cases. The (100)B orientation shows the largest VBO and CBO, equal to 0.53 eV and 2.44 eV, respectively. In contrast, the ($\bar{2}01$) orientation displays the smallest offsets, with a VBO of only -0.05 eV.  
The VBO reported for ($\bar{2}01$) aligns closely with those experimentally determined in 
Ref.~\cite{carey2017band} by means of XPS for films prepared by ALD. The discrepancy between the CBO shown in Fig.~\ref{fig:fig1} and the one reported Ref.~\cite{carey2017band} is due to the experimental value of the band gap (6.9 eV), as only the VBO is measured directly.
These results agree with the trend observed in previous modelling work.\cite{mu2020orientation} CBO for (010) and (001) orientations is almost identical to the values (2.24 eV and 1.86 eV, respectively) reported for strained slabs.\cite{mu2020orientation} This very close agreement is likely due to almost negligible, in-plane-only, atomic relaxation observed in slab geometries for the (010) orientation.\cite{mu_first-principles_2020} 
Notable differences between our results and those reported in Ref.~\onlinecite{mu2020orientation} are found for other orientations. These differences suggest that band offset calculations done using separate slab geometries are not always adequate, as they fail to model the chemical bonding at the interface between films and substrates. This failure is particularly evident when a non-negligible out-of-plane atomic displacement at the interface occurs,\cite{mu_first-principles_2020} which can substantially lower the electronic potential at the interface. We note that domain mixing among up to three competing strained variants, as recently predicted for heterostructures with monoclinic crystal systems,\cite{schubert2025strain} can provide an additional source of band offset broadening, at least for the ($h0l$) orientations.

\subsection{Pseudomorphic \ch{(Al_xGa_{1-x})2O3} films on $\beta$-\ch{Ga2O3} substrates}
The band gaps of strained \ch{(Al_xGa_{1-x})2O3} films pseudomorphically grown on $\beta$-\ch{Ga2O3} substrates have been calculated as a function of the Al content for each of the above-mentioned orientations. An increase in the Al content always yields a widening of the band gap, as expected, since the band gap of \ch{Al2O3} is larger than that of \ch{Ga2O3}. However, a non-negligible deviation from linearity is observed from a positive value of the bowing parameter, $b$, determined by fitting the relation:
\begin{equation}
E_g(x)=(1-x) E_g\left[\mathrm{G a_2 O_3}\right]+x E_g\left[\mathrm{A l_2 O_3}\right]-b x(1-x),
\end{equation}
where $E_g\left[\mathrm{G a_2 O_3}\right]$ and $E_g\left[\mathrm{A l_2 O_3}\right]$ are the band gaps of the pure compounds,  $\beta$-\ch{Ga2O3} and strained  $\theta$-\ch{Al2O3}.
The bowing parameters for \ch{(Al_xGa_{1-x})2O3} alloys are generally around 1 eV, in agreement with previously reported values.\cite{wang2018band} Only in the case of the [100] direction, a slightly higher value of 1.2 eV is observed.

\begin{figure*}[h]
\includegraphics[width=0.8\textwidth]{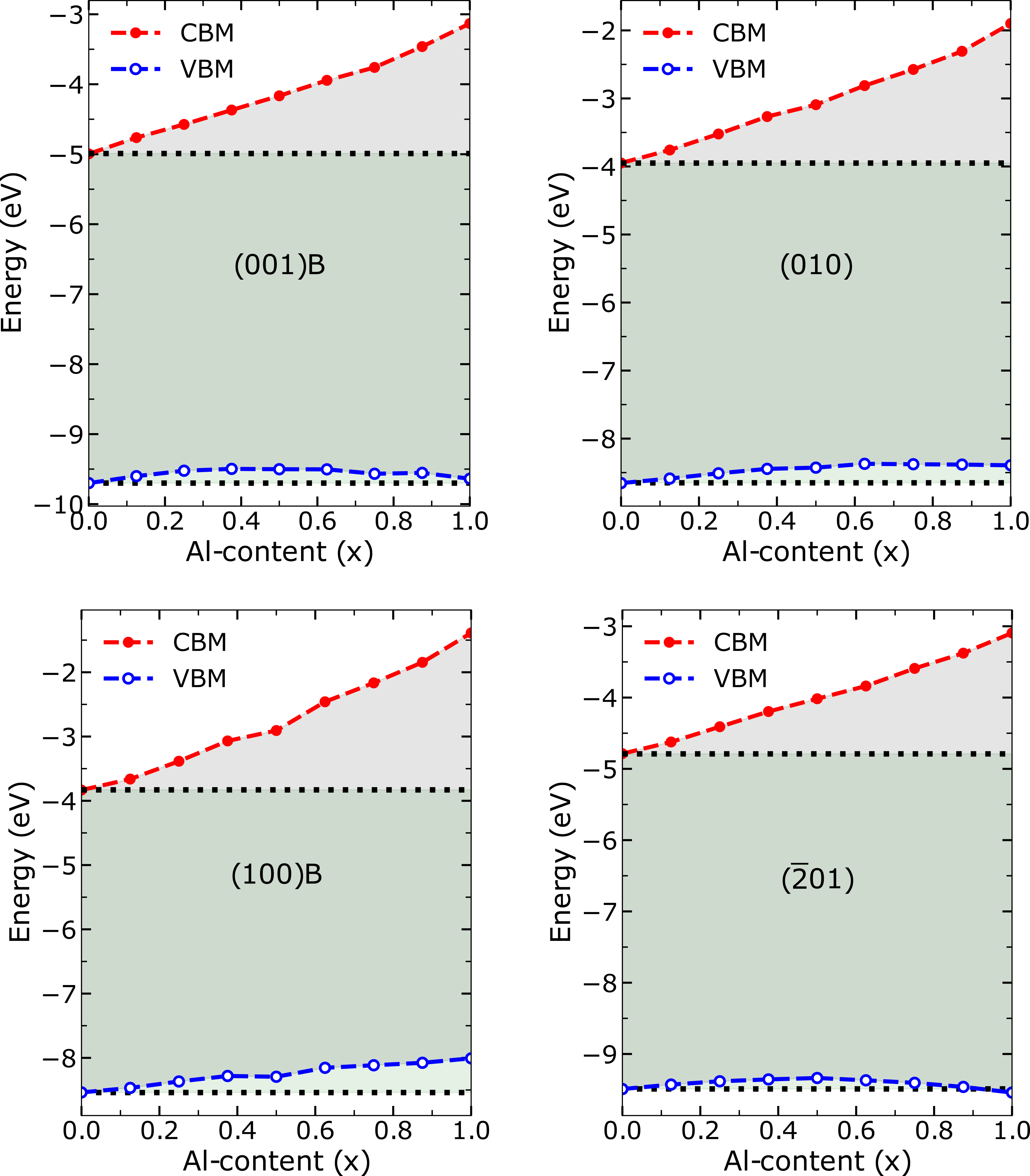}
\caption{The computed absolute energies of the conduction band minimum (CBM) and valence band maximum (VBM) (in eV) using superlattice structure calculations for $\beta$- \ch{(Al_xGa_{1-x})2O3}. The horizontal black dotted lines represent the CBM and VBM of $\beta$- \ch{Ga2O3}. The reported values are relative to the vacuum level, determined using a vacuum slab.}
\label{fig:fig2}
\end{figure*}

The knowledge of the absolute positions of the valence band maximum (VBM) and conduction band minimum (CBM) of the alloys is useful in view of applications, as these values are measured as ionisation energy and electron affinity.
To assess the absolute VBM and CBM, we conducted additional supercell calculations along the out-of-plane direction, with vacuum. We refer to this calculation as surface structures. The results are shown in Fig.~\ref{fig:fig2} for $\beta$-\ch{(Al_xGa_{1-x})2O3} alloys.

Each subplot shows the variation in CBM and VBM energies as a function of Al content, $x$, illustrating how band edge positions shift with increasing Al concentration. Absolute CBM and VBM of $\beta$-\ch{Ga2O3} are also shown for comparison. Notably, there is a significant orientation-dependent variation in the absolute CBM and VBM energies. For instance, the CBM for (001)B and ($\bar{2}01$) orientations lies at energies higher than for (010) and (100)B orientations, suggesting a strong influence of the film growth direction on the electron affinity. 

The observed trend suggests that Al substitution primarily affects the conduction band offset, increasing the electron affinity of the alloy without altering the position of the valence band significantly. This trend is likely due to the dominance of O 2p states at the VBM, which are less sensitive to the change in cation composition between Ga, and Al. This increase in CBO with Al content could enhance electron confinement in heterojunction structures, which is beneficial for device applications. Such orientation-dependent behaviour could be leveraged in device applications to tailor the band offset properties at heterojunction interfaces, depending on the desired electronic characteristics.


\newpage
\subsection{Schottky barrier diode modelling}
To investigate the practical effects of the \ch{Ga2O3}/\ch{(Al_xGa_{1-x})2O3} interface orientation on a device operation, we used the Ginestra® simulation software to reproduce the I--V and C--V characteristics of Pt/$\beta$-\ch{(Al_xGa_{1-x})2O3}/\ch{Ga2O3} SBD structures with 25\% Al content, using the calculated materials parameters for the four orientations $(\bar{2}01)$, $(010)$, $(100)$, and $(001)$.

\begin{figure*}
\includegraphics[width=1\textwidth]{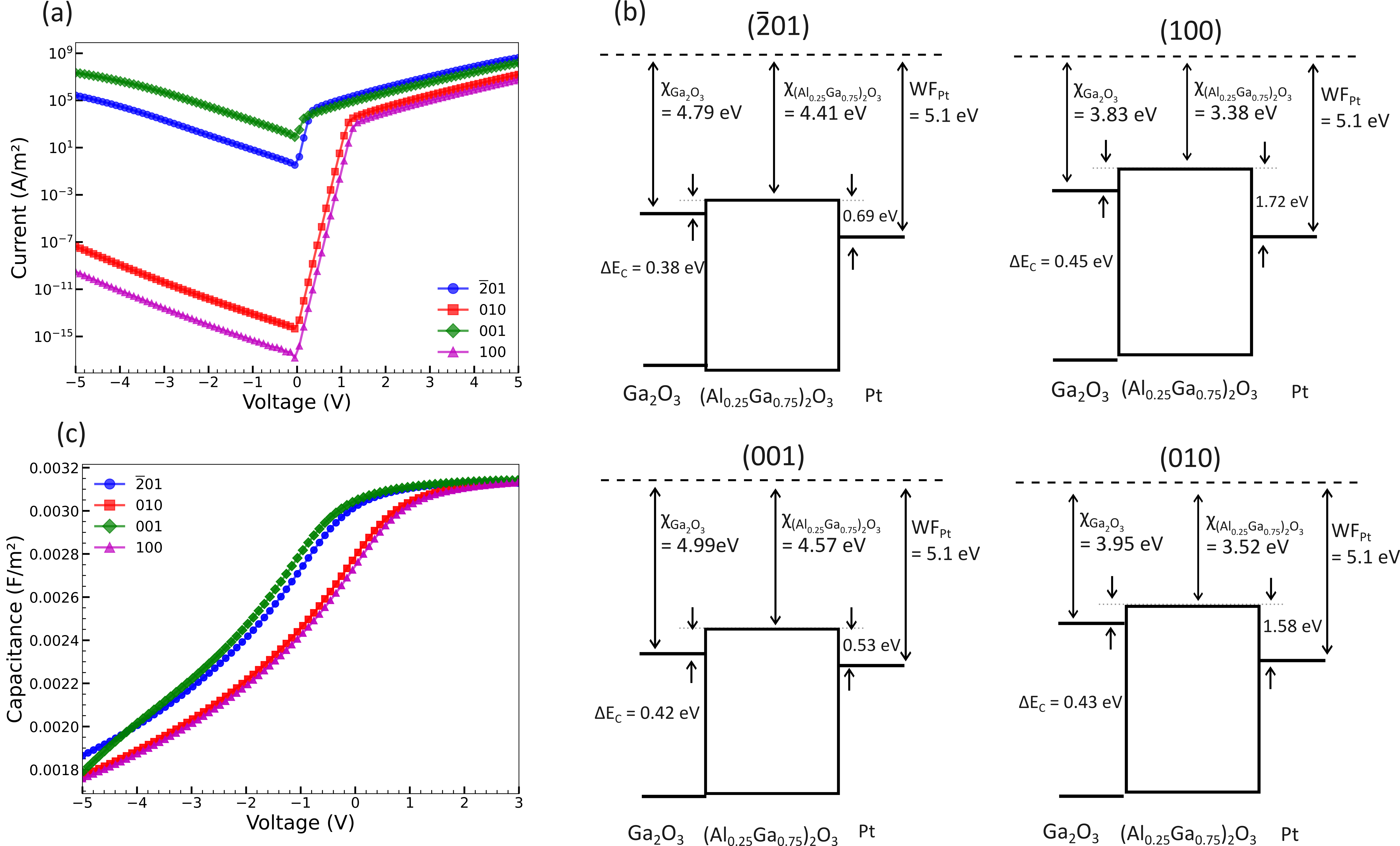}
\caption{(a) Simulated I--V characteristics of Pt/\ch{(Al_{0.25}Ga_{0.75})_2O_3}/\ch{Ga2O3} SBDs for different orientations: ($\bar{2}01$), (010), (001) and (100). The absolute value of the current is shown in the logarithmic vertical scale. (b) Energy band diagrams for the four orientations, illustrating work function (WF$_{Pt}$), electron affinity ($\chi$), and conduction band offset ($\Delta E_C$). (c) Simulated C--V characteristics for the same orientations, highlighting variations in capacitance with applied voltage. The \ch{(Al_{0.25}Ga_{0.75})_2O_3} relative dielectric permittivity is set to 11.15 and the thickness of the layer is set to 30 nm for all simulations.}
\label{fig:K_fig1}
\end{figure*}

As clearly visualized in the band diagrams of the simulated SBD [Fig.~\ref{fig:K_fig1}(b)], the interface orientation affects both the \ch{Ga2O3}/\ch{(Al_xGa_{1-x})2O3} conduction band offset---as expected due to the independent variation of the two materials' electron affinities---and the \ch{(Al_xGa_{1-x})2O3}/Pt  Schottky barrier height---the Pt work function being unaffected by the orientation.
Figure~\ref{fig:K_fig1} (a) shows the I--V characteristics of the simulated SBD for the four considered orientations. 
At negative bias (i.e. the reverse bias condition for the SBD) the charge transport is given by a combination of thermionic emission and tunnelling processes  \cite{latreche2019combined} from the Pt cathode at the Pt/$\beta$-\ch{(Al_xGa_{1-x})2O3} interface, with both the mechanisms strongly and inversely dependent on the Pt/$\beta$-\ch{(Al_{0.25}Ga_{0.75})_2O_3} Schottky barrier height (SBH). This is consistent with the simulation results shown in Fig. \ref{fig:K_fig1}, where a lower current magnitude corresponds to a higher Pt/$\beta$-\ch{(Al_{0.25}Ga_{0.75})_2O_3} SBH, and thus to a lower electron affinity of the specific \ch{(Al_{0.25}Ga_{0.75})_2O_3} orientation. Further Ginestra® simulations allowed us to inspect the specific nature of the negative bias charge transport by running different simulations while enabling a single transport mechanism at a time. For all four simulated \ch{(Al_{0.25}Ga_{0.75})_2O_3}/\ch{Ga2O3} interface orientations, the negative bias transport results were dominated by the tunnelling through the evolving trapezoidal/triangular \ch{(Al_{0.25}Ga_{0.75})_2O_3} barrier.

At positive bias (i.e. the forward bias condition for the SBD) the charge transport was also investigated by using Ginestra® and applying the same methodology used for the negative bias region. This results in a combination of diffusion current and thermionic emission at low bias, followed by a combination of thermionic emission and tunnelling processes at higher positive bias.
The contribution related to the diffusion process is recognizable in the low-voltage region, most noticeably for the (100) and (010) orientations, characterized by the exponential increase of the current with the ideal slope of $q/\ln(k_BT) = 60$ mV/decade at room temperature.
This originates from the presence of a depletion region in the \ch{Ga2O3} due to the built-in voltage $V_{bi}$ of the SBD (i.e. the difference between the \ch{Ga2O3} electron affinity and the Pt work function) imposed by the specific orientation. Noticeably, the presence of a depletion region barrier in the \ch{Ga2O3} is the main bottleneck for the majority carriers’ transport at low field. Upon the application of a positive bias the barrier is progressively lowered, leading to an exponential increase of the current due to majority carriers’ diffusion over the \ch{Ga2O3} barrier.

As the \ch{Ga2O3} reaches flat-band ($V = V_{bi}$) or accumulation condition ($V > V_{bi}$), with no depletion region barrier in place, the charge transport becomes once again limited by the thermionic emission and  tunnelling through an evolving trapezoidal/triangular \ch{(Al_{0.25}Ga_{0.75})_2O_3} barrier.

The voltage dependency of the limiting transport mechanism is clearly visualized in the I--V characteristics of Fig. \ref{fig:K_fig1} (a): at low-voltage, the current is characterized by an exponential increase, transitioning into a more complex trend at high voltage. In the low voltage regime, the current magnitude is governed by the \ch{Ga2O3} electron affinity, while at high voltage it is determined by the \ch{(Al_{0.25}Ga_{0.75})_2O_3}/\ch{Ga2O3} conduction band offset.

From the presented analysis, we note that the specific \ch{(Al_{0.25}Ga_{0.75})_2O_3}/\ch{Ga2O3} orientation can be used for tailoring the performance of the devices. For example, using the $(100)$ orientation results in an SBD with an extremely high $I_{ON}/I_{OFF}$ ratio, while the $(\bar{2}01)$ orientation results in an SBD characterized by a low built-in voltage (thus fast turn on) and higher forward current. 

The simulated high-frequency (1 MHz) C--V characteristics of the same SBD are shown in Fig.~\ref{fig:K_fig1} (c), exhibiting very similar behaviours regardless of the specific orientation. The simulated C--V characteristics resemble those of a typical MOS capacitor, with a deep-depletion trend at negative voltage and with the maximum capacitance at positive bias voltage (i.e. in nominal accumulation) determined by the \ch{(Al_{0.25}Ga_{0.75})_2O_3} layer dielectric constant and thickness. The only significant difference between the simulated C--V characteristics is a voltage shift given by the orientation-dependent built-in voltage.

The simulated C--V characteristics are remarkably similar to those of previously reported for \ch{(Al_{0.08}Ga_{0.92})_2O_3}/\ch{Ga2O3}  modulation doped field effect transistors,\cite{okumura_demonstration_2019}  presenting a fundamental MOS structure very similar to the ones presented and simulated in this work. It is worth noting that the reported \ch{(Al_{0.08}Ga_{0.92})_2O_3}/\ch{Ga2O3}  structure is capable of confining a two-dimensional electron gas at the interface and, given the similarities with our simulated structures, we can speculate that they can also exhibit such a capability.

\begin{figure}[ht]
\includegraphics[width=0.8\textwidth]{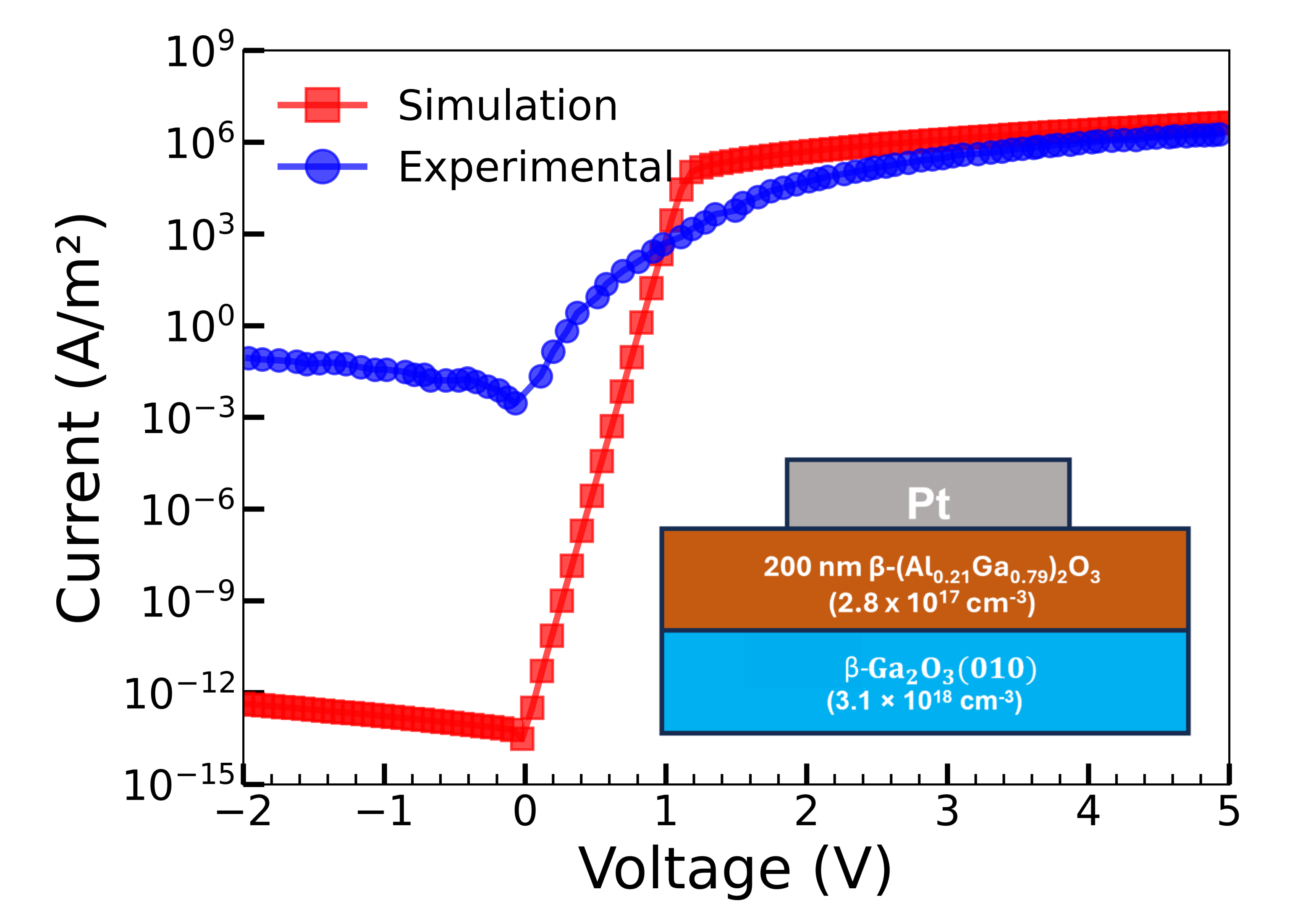}
\caption{Comparison of simulated (ideal) and experimental I--V data from Ref.~\citenum{mudiyanselage_ultrawide_2023} for \ch{(Al_{0.21}Ga_{0.79})_2O_3}/\ch{Ga2O3}(010) SBD. The thickness of the \ch{(Al_{0.21}Ga_{0.79})_2O_3} layer (200 nm) in the simulation is identical to the \ch{(Al_{0.21}Ga_{0.79})_2O_3} thickness used in device of Ref. \citenum{mudiyanselage_ultrawide_2023}. The electron affinity values (3.60 eV for \ch{(Al_{0.21}Ga_{0.79})_2O_3} and 3.95 eV for \ch{Ga2O3} (010)) was interpolated from table S2. The inset shows a schematic of the simulated device with the doping concentrations for each layer.
}
\label{fig:K_fig3}
\end{figure}

The experimental I--V measurements provided in Ref.~\cite{p_sundaram_-alxga1x2o3ga2o3_2022,mudiyanselage_ultrawide_2023} were used to validate our simulations.
In particular, we want to assess how accurately device simulations based on band offsets and dielectric constants from first principles can match published experimental I--V data. Values of the band offset and electron affinity (3.60 eV) for 21\% Al content are obtained by interpolation of the first principles results shown in Table S2. These values are used in the device simulations to generate I--V characteristics for comparison against the  experimental results reported by Mudiyanselage \emph{et al.}\cite{mudiyanselage_ultrawide_2023}. The simulation is performed for an ideal  Pt/\ch{(Al_{0.21}Ga_{0.79})_2O_3}/\ch{Ga2O3} structure, in absence of  any defect profiles. The thickness of the \ch{(Al_{0.21}Ga_{0.79})_2O_3} layer is set to 200 nm for the interface oriented in the (010) direction. Additional simulation parameters are listed in table S1. The simulated I--V together with the experimental I--V are plotted in Fig.~\ref{fig:K_fig3}. 

The trend is well reproduced, with discrepancies in the current magnitude. At low positive bias, the experimental curve deviates from the simulated characteristic, likely due to the presence of defects at the \ch{(Al_{0.21}Ga_{0.79})_2O_3}/\ch{Ga2O3} interface, which are not accounted for in our simulations. Interface defects are known to induce higher current through Schockley-Read-Hall carrier generation/recombination mechanisms. 

The experimental forward current does not present a sharp transition from the diffusion to the tunnelling regime, possibly indicating a heterojunction potential less abrupt than the simulated one, or a surface potential pinning effect due to the presence of interface states at the \ch{Ga2O3}/\ch{(Al_{0.21}Ga_{0.79})_2O_3} interface. The discrepancy between the experimental and simulated I--V  in negative bias region is likely caused by defects in the bulk of the \ch{(Al_{0.21}Ga_{0.79})_2O_3} layer. 

\begin{figure}[h]
\centering
\includegraphics[width=0.9\textwidth]{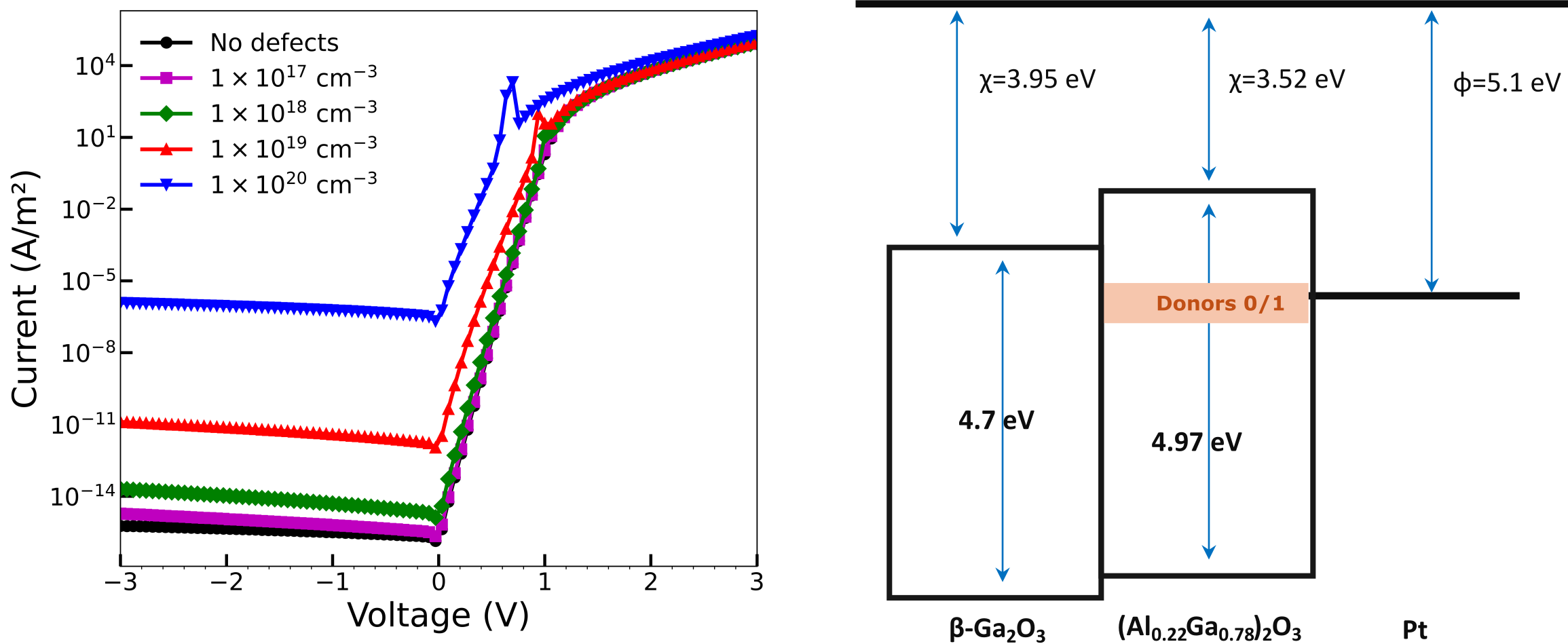}
\caption{ Simulated I--V curves for \ch{(Al_{0.22}Ga_{0.78})_2O_3}/\ch{Ga2O3} (010) SBD considering different defect concentrations in the \ch{(Al_xGa_{1-x})2O3} layer. Direct and trap-assisted tunnelling conduction mechanisms are included in the simulation.
}
\label{fig:s-fig5}
\end{figure}

\begin{figure}[ht]
\includegraphics[width=0.8\textwidth]{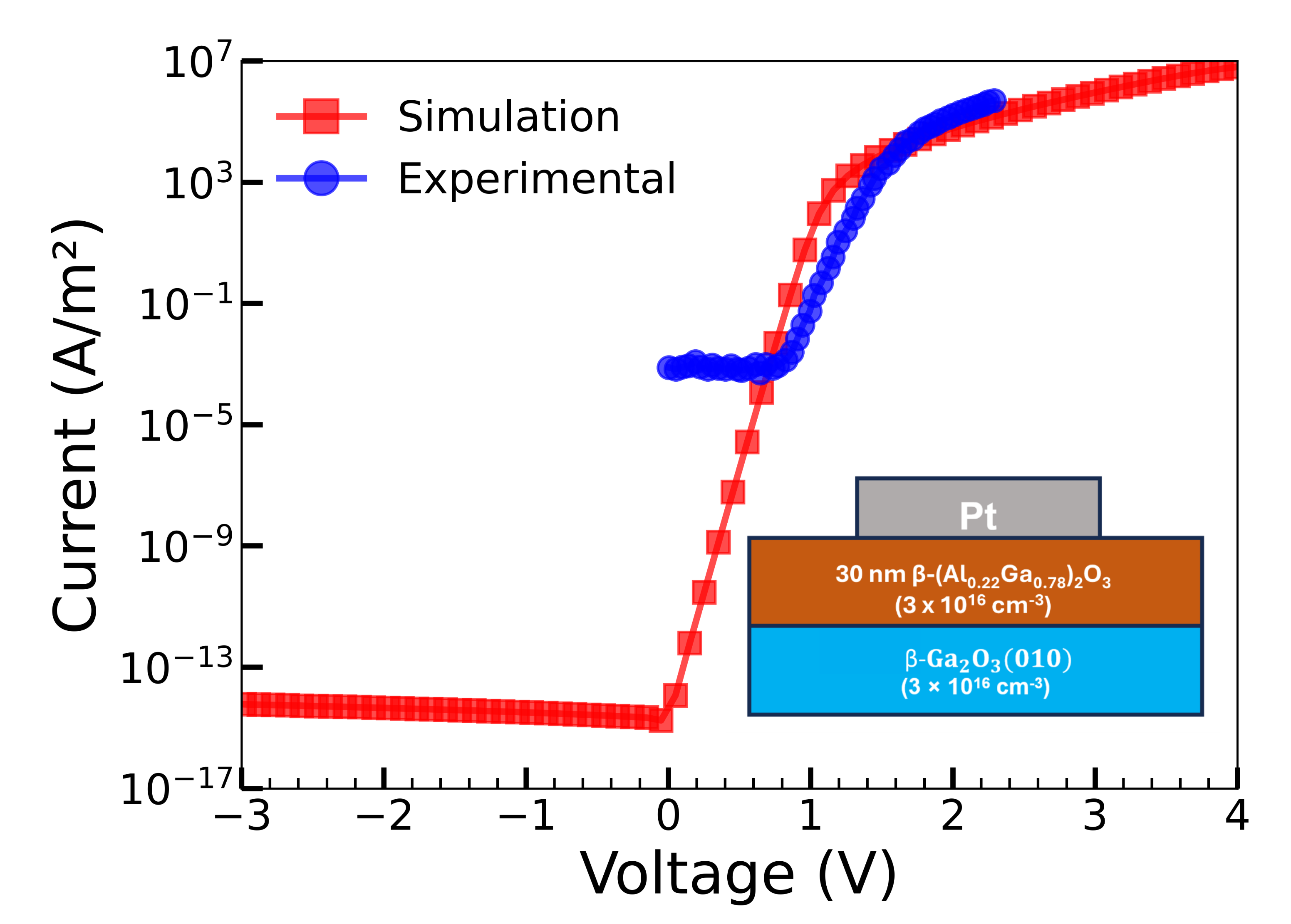}
\caption {Comparison of simulated (ideal) and experimental I--V data from Ref.\citenum{p_sundaram_-alxga1x2o3ga2o3_2022} for \ch{(Al_{0.22}Ga_{0.78})_2O_3}/\ch{Ga2O3}(010) SBD. The interpolated electron affinity value of \ch{(Al_{0.22}Ga_{0.78})_2O_3}(010) is 3.58 eV and \ch{Ga2O3}(010) is 3.95 eV (table S2). The \ch{(Al_{0.22}Ga_{0.78})_2O_3} thickness and mobility values considered for the simulation are 30 nm and 184 $\ \text{cm}^2\cdot\text{V}^{-1}\cdot\text{s}^{-1}$, respectively (identical to the values provided by Ref. \citenum{p_sundaram_-alxga1x2o3ga2o3_2022}). The inset shows a schematic of the simulated device with the doping concentrations for each layer.
}
\label{fig:K_fig4}
\end{figure}

The first principles calculated band offset parameters were used to simulate SBD I-V characteristics and compare them to previously published experimental I-Vs (figures \ref{fig:K_fig3} and \ref{fig:K_fig4}).\cite{mudiyanselage_ultrawide_2023,p_sundaram_-alxga1x2o3ga2o3_2022} The simulations matched remarkably well the experimental data in a bias regime where the current is most sensitive to the band offsets values, hence validating the first principles calculations. Matching the experimental I--Vs across the full bias range is outside the scope of this work. However, by considering a defect band in the \ch{(Al_{x}Ga_{1-x})_2O_3} layer (donor defect 1.38 eV below the \ch{(Al_{x}Ga_{1-x})_2O_3} conduction band) aligned with the Pt Fermi level it is possible to clearly illustrate how the SBD current at reverse bias is affected by the presence of defects (figure~\ref{fig:s-fig5}).

Figure~\ref{fig:K_fig4} shows the experimental I--V characteristics reported by Sundaram \emph{et al.}.\cite{p_sundaram_-alxga1x2o3ga2o3_2022}, along with the simulated I–V curves generated in this study. The I--V curves are obtained for a SBD fabricated on MOCVD-grown \ch{(Al_{0.22}Ga_{0.78})_2O_3} on $\beta$-\ch{Ga2O3}(010) substrate. The doping concentration in the $\beta$-\ch{Ga2O3} substrate and in the 30 nm \ch{(Al_{0.22}Ga_{0.78})_2O_3} layer are assumed to be $3 \times 10^{16}$  $\text{cm}^{-3}$, identical to the doping level reported in Ref.~\citenum{p_sundaram_-alxga1x2o3ga2o3_2022}. The differences in the I–V characteristics between Figs.~\ref{fig:K_fig3} and ~\ref{fig:K_fig4} arise from the varying \ch{(Al_xGa_{1-x})2O3} layer thicknesses and doping concentrations (See Fig.~S5). 

In the forward bias region, Fig.~\ref{fig:K_fig4} shows a very good agreement between the simulation and the published experimental data. As for the simulation reported above, the simulated I--V curve is obtained for an ideal case, in absence of any defect profile in the structure. It is interesting to note that in the forward region the experimental I--V curve shows a voltage dependence which follows the ideal curve more closely than in the previous case, reported in  fig.~\ref{fig:K_fig3}). We also note that the I--V profile present a sharper transition from diffusion to thermionic regimes, suggesting a more abrupt \ch{(Al_{0.22}Ga_{0.78})_2O_3}/\ch{Ga2O3} interface for the device of Ref.~\cite{p_sundaram_-alxga1x2o3ga2o3_2022} than that of Ref.~\cite{mudiyanselage_ultrawide_2023}.  The main differences between the experimental I--V curve and the simulated data observed in fig.~\ref{fig:K_fig4} are the horizontal shift, possibly due to a mismatch in the built-in voltage or the presence of negative charge in the materials' stack.  
As previously discussed, the positive shift in the experimental I--V could be caused by either i) an experimental Schottky barrier larger than the ideal value (\emph{i.e.} the difference between the Pt work function and the \ch{Ga2O3} Fermi level) considered in the simulations or ii) the presence of negative charge and depletion at the \ch{(Al_{0.22}Ga_{0.78})_2O_3}/\ch{Ga2O3} interface. 
The higher reverse current in the negative bias region has probably a similar origin as the previous example caused by defects which increase the Schottky reverse leakage current.

\section{Conclusion}
We have conducted computational modeling using the Heyd-Scuseria-Ernzerhof hybrid functional and Technology Computer-Aided Design models to investigate band offset, I--V and C--V for different growth orientations and alloy compositions in \ch{Ga2O3}/\ch{(Al_xGa_{1-x})2O3} 
heterojunctions. Strain from lattice mismatch affects both VBM and CBM, with distinct behaviours observed across ($\bar{2}01$), (010), (001) and (100) orientations due to the high anisotropy of the monoclinic phase. We examined the effect of the strain on band offset by incorporating its effect on bulk energy levels and the potential lineup at the interface to ensure a consistent reference. The high sensitivity to growth orientation and strain may justify the discrepancies in previously published results. Additionally, we report C--V and I--V simulations for \ch{Ga2O3}/\ch{(Al_{0.25}Ga_{0.75})_2O_3}/Pt SBD across the four growth orientations, revealing significant orientation dependence given the different materials' electron affinities and the resulting conduction band offsets. To validate our findings, we simulated and compared with experimental I--V results for Al concentrations of 22 \% and 21 \%, achieving reasonable agreement in the forward bias. These results provide insights into the device structure and the potential defect and highlight the critical role of orientation and strain engineering in optimising wide-bandgap devices for power electronics.
\section{Associated content}
Detailed information on the modelling parameters, including the structures used for simulations in both bulk and interface calculations, as well as a comprehensive list of parameters employed in the device simulation, is given in the Supplemental Material~\cite{hi}.
\begin{acknowledgments}
This work was supported by a research grant from the Department for the Economy Northern Ireland (DfE) under the US-Ireland R\&D Partnership Programme (USI 195). The Tyndall National Institute is supported by Research Ireland under the US-Ireland Research and Development Partnership (21/US/3755).
Access to the computing facilities and support from the Northern Ireland High-Performance Computing (NI-HPC) service funded by EPSRC (EP/T022175), and to the UK national high-performance computing service, ARCHER2, through the UKCP consortium and funded by EPSRC (EP/X035891/1) are also gratefully acknowledged.
\end{acknowledgments}
\section*{Data Availability Statement}
Computational parameters are provided in the main manuscript and Supplemental Material~\cite{hi}. Additional data are available from the corresponding author upon reasonable request.
\bibliography{reference}

\end{document}